# Observation of modulation instability and rogue breathers on stationary periodic waves

Gang Xu[1], Amin Chabchoub[2,3], Dmitry E. Pelinovsky[4,5], Bertrand Kibler[1]

[1]*Laboratoire Interdisciplinaire Carnot de Bourgogne, UMR6303 CNRS-UBFC, 21000 Dijon, France*
[2]*Centre for Wind, Waves and Water, School of Civil Engineering, The University of Sydney, Sydney, NSW 2006, Australia*
[3]*Marine Studies Institute, The University of Sydney, Sydney, NSW 2006, Australia*
[4]*Department of Mathematics, McMaster University, Hamilton, Ontario, Canada, L8S 4K1*
[5]*Institute of Applied Physics RAS, Nizhny Novgorod, 603950, Russia*

We report an experimental study on the modulation instability process and associated rogue breathers for the case of stationary periodic background waves, namely dnoidal and cnoidal envelopes. Despite being well-known solutions of the nonlinear Schrödinger equation (NLSE), the stability of such background waves has remained unexplored experimentally until now, unlike the constant-amplitude plane wave. By means of two experimental setups, namely, in nonlinear optics and hydrodynamics, we observe the spontaneous modulation instability gain seeded by input random noise and the formation of rogue breathers induced by a coherent perturbation. Our observations are in excellent agreement with the NLSE dynamics.

*Introduction.* During the last decades, the modulation instability (MI) phenomenon has attracted a significant research interest in a variety of nearly conservative wave systems (water surface, plasmas, guided laser light, electrical transmission lines, and Bose-Einstein condensates) described by the nonlinear Schrödinger equation (NLSE) in its many forms [1-12]. This includes the linear stability analysis of the plane waves and the subsequent nonlinear stage of MI, namely the formation of localized waves such as solitons and breathers, as well as multi-breather complexes. Note that the space-time evolution of the nonlinear stage of MI strongly depends on the input perturbation considered. In particular, a broad class of modulationally unstable initial conditions can be described by so-called continuous spectrum solutions of the NLSE [13-15]. Beyond the plane waves, within the class of stationary solutions of the focusing NLSE, a wide range of periodic solutions known as dnoidal (dn) and cnoidal (cn) waves are also modulationally unstable against small perturbations [16,17]. Note that the plane wave is just a limiting case and thus a special case of dn-periodic waves. These stationary periodic waves are highly relevant in the studies of extreme wave formation and their generalization, resulting from MI in more practical wave conditions [18-20] and from the development of integrable turbulence [21].

Although the mathematical description of MI for such stationary periodic waves is well understood [22], no experimental observation of the phenomenon has been reported so far. Propagation of the stationary periodic envelopes has been conducted in distinct water wave facilities [23-24], but without reporting on the stability against small perturbations. For nonlinear optical studies, we can only mention the experimental evidence of the cnoidal wave self-compression in a photorefractive crystal [25]. By contrast, cnoidal waves have been widely studied in the framework of the Korteweg-de Vries equation, which is a shallow water framework, and related physical systems [2-3].

In this work, we provide an overview of both the noise-driven and the coherent seeding regimes of MI for stationary periodic waves of the NLSE. To investigate such regimes on a relevant range of parameters, we provide two complementary experimental setups based on light-wave propagation in an optical fiber and wave propagation in a water wave tank. Despite completely different timescales of the nonlinear dynamics are involved, the MI and rogue breathers are described by the same theoretical foundation of the universal NLSE. We quantitatively confirm the spontaneous MI gain and the formation of rogue breathers on a stationary periodic background. It is worth to highlight that quantitative generation of temporal *cn*- or *dn*-wave solutions has never been tested in optics until now since it remains a challenging task.

In optical experiments, we show that the family of cn-waves is more robust against noise than dn-waves, thus being of potential interest for optical data processing and transmission [26]. This is related to weak MI gains of the cn-waves. As a result, we only observe the MI and rogue breathers on the dn-wave in the optical fibers. On the other hand, in hydrodynamical experiments, we observe rogue breathers on both backgrounds of the dn-waves and the cn-waves, however, observation of MI gains from random noise is difficult due to the limitations imposed by the length of the wave facility.

*Theoretical description.* Our theoretical framework is based on the dimensionless form of the universal self-focusing 1D-NLSE,

$$i\psi_\xi + \tfrac{1}{2}\psi_{\tau\tau} + |\psi|^2\psi = 0 \qquad (1)$$

where subscripts stand for partial differentiations. Here $\psi$ is a wave envelope which is a function of $\xi$ (a scaled propagation distance) and $\tau$ (a co-moving time with the wave-group velocity). The NLSE has exact complex breather solutions as well as simpler ($\xi$-) stationary ($\tau$-) periodic solutions of the dnoidal and cnoidal type, expressed in terms of elliptic functions [27].

The positive-definite dn-periodic waves and the sign-indefinite cn-periodic waves are respectively given by

$$\psi_{\text{dn}} = dn(\tau, k) e^{i(1-k^2/2)\xi} \quad (2)$$

and

$$\psi_{\text{cn}} = k\, cn(\tau, k) e^{i(k^2-1/2)\xi} \quad (3)$$

where $k$ is the modulus of elliptic functions ($0 \leq k \leq 1$), which gives the period of the wave function $T_c = 2K(k)$ for (2) and $T_c = 4K(k)$ for (3), where $K(k)$ is the complete elliptic integral. One can then obtain the angular frequency interval of the corresponding comb spectra $\Omega_c = 2\pi/T_c$ (see illustrations in Fig. 1).

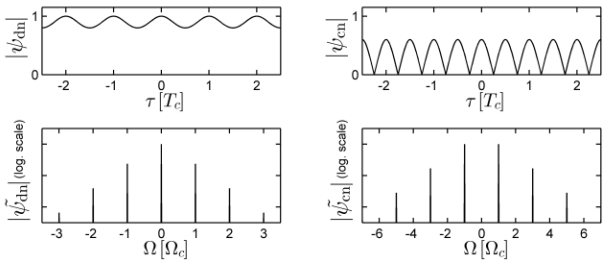

FIG. 1. Examples of (left) dn- and (right) cn-waves for $k = 0.6$. *Top panels:* Temporal profiles. *Bottom panels:* Spectral profiles.

For $k \to 1$, both periodic wave families converge to the envelope soliton (sech-shape). For $k \to 0$, the dn-wave and the cn-wave tend to the plane waves with either normalized or vanishing amplitude, respectively. From Fig. 1, it is worth noting that cn-waves are characterized by a spectral envelope mainly driven by bi-chromatic waves, whereas dn-waves correspond to modulated single-frequency backgrounds [24]. Recall that the above waves belong to the restricted group of stationary periodic waves with trivial phase. More general elliptic wave solutions with nontrivial phase can be also analysed [17,22].

The interaction between dispersive and nonlinear effects leads to MI phenomenon for the plane wave in the presence of noise (spontaneous regime) or a weak frequency-shifted signal wave (induced regime) [1-5]. The linear stability analysis of periodic waves was also elaborated in detail (see for instance Ref. [17]). It was found that both dn- and cn-waves are modulationaly unstable with respect to long-wave perturbations. We provide below the forms of MI growth rate according to parameters of the periodic waves.

The general evolution of an initial perturbation onto $\psi_{dn}$ or $\psi_{cn}$ can be expressed as $e^{\Gamma \xi}$, where the MI growth rate (in amplitude) is mainly given by the real part of $\Gamma = \pm 2i\sqrt{P(\lambda)}$, where $P(\lambda)$ can be calculated by using the following relations for dn- and cn-waves:

$$P_{dn} = \lambda^4 - \left(1 - \frac{k^2}{2}\right)\lambda^2 + \frac{k^4}{16} \quad (4)$$

and

$$P_{cn} = \lambda^4 - \left(k^2 - \frac{1}{2}\right)\lambda^2 + \frac{1}{16} \quad (5)$$

Here, $\lambda$ is the spectral parameter defined in the Lax spectrum of the Zakharov-Shabat spectral problem (see [22] for details). The numerical scheme of computing the eigenvalues is based on the discretization of the frequency comb interval [22]. We here relate eigenvalues in the Lax spectrum to parameters of the periodic waves and the frequency range of perturbations that can be investigated in each frequency interval ($0 \leq |\Omega| \leq \Omega_c$). The instability arises only if $\lambda$ belongs to the bands of the Lax spectrum with $Re\{\lambda\} \neq 0$.

Figure 2 shows calculations of MI growth rate for both dn- and cn-periodic waves as a function of normalized angular frequencies of perturbation (i.e., a practical picture for experimental studies) and for distinct values of their governing parameter (i.e., the modulus of elliptic functions). In the case of dn-waves evolving from $k = 0$ towards 1 (see Fig. 2a), the MI starts from the plane wave limit, where it occurs for frequencies $0 \leq |\Omega| \leq 2$ ($\Omega_c = 2$) as well as characterized by a maximum growth rate equal to 1 at $|\Omega| = \sqrt{2}$. Then, when $k$ increases, the MI spectral bandwidth (here equivalent to $|\Omega_c|$) continuously reduces as the wave period $T_c$ increases. At the same time, $Re\{\Gamma\}$ also decreases and vanishes in the limit $k \to 1$ of the stable solitons. On the contrary, for cn-waves evolving from $k = 0$ (i.e., the zero background limit: no MI) towards 1 (see Fig. 2b), the MI starts to grow near $|\Omega| = \Omega_c = 1$ and then MI bands enlarge until reaching maximal growth rate and bandwidth for $k \sim 0.83$. This maximum observed at $\sim 0.36\, \Omega_c$ remains significantly smaller than the growth rates obtained for dn-periodic waves. After that, MI growth rate decreases for higher values of $k$ since the wave period $T_c$ increases and cn-waves tend to the stable sech-solution ($k = 1$).

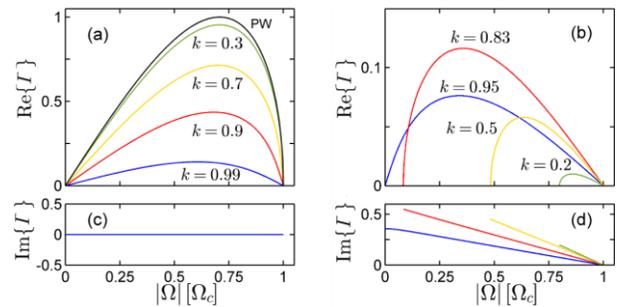

FIG. 2. Calculated MI growth rate $Re\{\Gamma\}$ as a function of normalized angular frequencies of perturbation for (a) dn-waves and (b) cn-waves. (c,d) Corresponding calculated $Im\{\Gamma\}$. The plane-wave limit (PW) is plotted with a black line in panel (a).

Unlike dn-waves, an original specificity of the MI phenomenon in the case of cn-waves is that the imaginary part of $\Gamma$ is nonzero (see Figs. 2c-2d), thus leading to oscillations of the genuine growth rate along propagation distance. It means that the fastest growing perturbation changes according to Im$\{\Gamma\}$. Consequently, the MI growth rate indicated in Fig. 2(b) does not report the fine $\xi$-dependent spectral structure, but only the asymptotic solution for very large distances.

MI can be triggered by noise when the wave field evolves over a significant propagation distance as generally the case in optical experiments. To address this, we provide in Fig. 3 a typical example of spontaneous modulation instability (MI) gain bands that emerge in the case of cn-periodic waves (here, $k = 0.5$). Figure 3(a) reports the evolution of power spectra as a function of propagation distance obtained from numerical simulations of the NLSE. The simulated output spectra result from an averaging over 100 simulations based on a small noise with random spectral phase superposed to the initial cn-wave. We found that the exponential growth of initial noise exhibits spectral oscillations, as well as distinct fastest growing frequencies along propagation distance. This confirms the impact of the nonzero imaginary part of $\Gamma$.

We compared the corresponding accumulated MI gain obtained for various propagation distances to the analytical predictions (see Fig. 3(b)). The accumulated MI gain differs in bandwidth and shapes for distinct propagation distances. In the first steps of propagation (bottom panel), a first spectral band emerges and then continuously drifts to larger frequency de-tunings. By increasing the propagation distance (middle and top panels), more spectral bands appear and fill the frequency interval defined by Re$\{\Gamma\}$ (dotted lines) while the overall gain also increases. In order to describe the fine $\xi$-dependent MI spectral structure, Im$\{\Gamma\}$ has to be carefully included in the theoretical predictions, whereas Re$\{\Gamma\}$ only gives the asymptotic solution for very large distances.

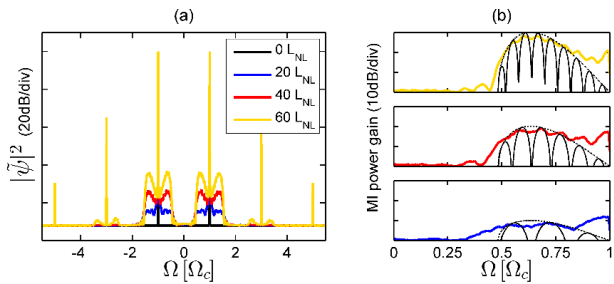

FIG. 3. (a) Power spectra obtained from NLSE simulations showing the spontaneous emergence of MI gain bands for the case of a cn-periodic wave ($k = 0.5$) as a function of propagation distance. (b) Corresponding accumulated MI power gain deduced from panel (a) and compared to theoretical predictions (black lines) for the cn-wave studied. Solid (dotted) lines are calculated with (without) Im$\{\Gamma\}$. Panels from bottom to top correspond to increasing propagation distances.

For both cases of periodic waves, it was shown in [19-20,22] that the roots of the polynomial $P(\lambda)$ for which $\Gamma = 0$ can be used to construct the rogue breather (or rogue wave, RW) solutions $\psi^{RW}$ on the corresponding periodic background. Such solutions generalize the well-known Peregrine's breather (or rogue wave) on the continuous wave background. The analytical expression of RW on the dn-periodic wave can be written as:

$$\psi_{dn}^{RW}(\tau,\xi) = \left[dn(\tau,k) + \frac{F(\tau,\xi)}{G(\tau,\xi)}\right] e^{i(1-k^2/2)\xi} \quad (6)$$

where

$$F(\tau,\xi) = [1 - 2i\,\text{Im}\{\theta(\tau,\xi)\} - |\theta(\tau,\xi)|^2][dn(\tau,k)^2 + \sqrt{1-k^2}]$$

$$G(\tau,\xi) = [|\theta(\tau,\xi)|^2 + 1]dn(\tau,k) + 2\left(1 - \sqrt{1-k^2}\right)\text{Re}\{\theta(\tau,\xi)\}\,sn(\tau,k)\,cn(\tau,k)$$

and

$$\theta(\tau,\xi) = \left[dn(\tau,k)^2 + \sqrt{1-k^2}\right] \times \left[-2(1+\sqrt{1-k^2})\int_0^\tau \frac{dn(\tau',k)^2}{[dn(\tau',k)^2+\sqrt{1-k^2}]^2}d\tau' - i\xi\right].$$

This RW tends to the Peregrine's breather when $k \to 0$, whereas it looks like a two-soliton interaction for $k \to 1$ [19]. The maximum amplitude of the RW is equal to $M = 2 + \sqrt{1-k^2}$ at $(\tau,\xi) = (0,0)$, which is the magnification factor of the RW.

Similarly, the analytical expression of RW on the cn-periodic wave can be written as:

$$\psi_{cn}^{RW}(\tau,\xi) = \left[k\,cn(\tau;k) + \frac{F(\tau,\xi)}{G(\tau,\xi)}\right] e^{i(k^2-1/2)\xi} \quad (7)$$

where

$$F(\tau,\xi) = k[1 - 2i\,\text{Im}\{\theta(\tau,\xi)\} - |\theta(\tau,\xi)|^2] \times [cn(\tau,k)\,dn(\tau,k) + i\sqrt{1-k^2}\,sn(\tau,k)]$$

$$G(\tau,\xi) = [|\theta(\tau,\xi)|^2 + 1]dn(\tau,k) + 2\text{Re}\{\theta(\tau,\xi)\}\,k\,sn(\tau,k)\,cn(\tau,k)$$

and

$$\theta(\tau,\xi) = \left[k^2 cn(\tau,k)^2 + ik\sqrt{1-k^2}\right] \times \left[-2(k + i\sqrt{1-k^2})\int_0^\tau \frac{k^2 cn(\tau',k)^2}{[k^2 cn(\tau',k)^2+ik\sqrt{1-k^2}]^2}d\tau' - i\xi\right].$$

This RW looks like a propagating soliton for $k \to 0$, whereas it can be compared to a two-soliton interaction for $k \to 1$ [19]. Again, the maximum occurs at $(\tau,\xi) = (0,0)$ and is equal to $2k$ which gives the $k$-independent magnification factor $M = 2$.

Typical illustrations of the above RW solutions and described space-time dynamics are depicted in Fig. 4, for different values of $k$. We confirm that the maximal amplitude of RW on the cn-periodic waves is always lower than that on the dn-periodic wave.

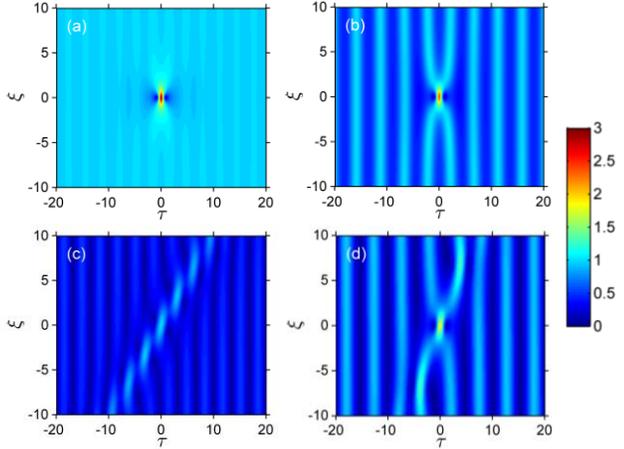

FIG.4. Theoretical space-time dynamics of RW solutions $|\psi^{RW}(\tau,\xi)|$ on periodic dn- (top panels) and cn- (bottom panels) periodic waves, namely (a-b) RW on the dn-periodic waves for $k = 0.3, 0.9$, respectively. (c-d) RW on the cn-periodic waves for $k = 0.5, 0.95$, respectively.

*Experimental setups.* The observation of spontaneous MI gain and RW solutions is different in nonlinear optics and hydrodynamics due to their restricted ranges of parameters, since all the experimental parameters (e.g., wave period and amplitude, fiber dispersion and nonlinearity) are embedded into a single parameter of the stationary periodic wave, namely the modulus $k$ of elliptic functions. To find the correspondence between theory and experimental parameters, we refer the reader to the normalization relations given in the following.

The spontaneous MI that grows from small random noise requires a long propagation length with almost no dissipation because the expected MI gain is lower for the stationary periodic waves than for the plane wave. Nonlinear fiber optics represents a suitable solution for this issue and provide a direct spectral characterization of the MI dynamics. By contrast, rogue breathers on both the dn- and cn-periodic waves can be observed in the water wave tank, whereas only the RW on the dn-periodic wave can be generated with light waves since their precise arbitrary waveform shaping at various periodicities is far more difficult. We emphasize that all our experiments are designed in such a way as to prevent as much as possible any contribution from higher-order effects beyond the standard focusing NLSE, but losses can still affect our results. Non-ideal conditions would induce the emergence of asymmetric wave profiles or/and introduce some complex spatial recurrence phenomenon, thus diverging from the NLSE dynamics during propagation [28-31].

Our experimental setups (depicted in Fig. 5) are based on the propagation of arbitrarily shaped light waves in optical fibers and a common water wave tank. Each system is capable of synthesizing nontrivial exact periodic wave profiles in the temporal domain, i.e., a prerequisite for confirming the existence of their genuine instability.

For light waves, the initial state is obtained through the optical pulse shaping with phase and amplitude controls in the spectral domains. This specific processing of a home-made optical frequency comb source allows to generate exact wave profiles with a specific period fixed by the frequency spacing of the optical comb. Nonlinear propagation is then studied in different lengths of the same standard single-mode fiber (SMF28) by an appropriate choice of the input average power. At fiber output, the power profiles are characterized in both time and frequency domains by means of an ultrafast optical sampling oscilloscope (with sub-picosecond resolution) and an optical spectrum analyzer (with 2.5 GHz resolution). We invite the reader to refer to Refs. [28,32-33] for more details about the system characterization. In general, only slight discrepancies can be noticed mainly ascribed to the linear propagation losses in our optical fiber and some artefacts of the initial wave shaping.

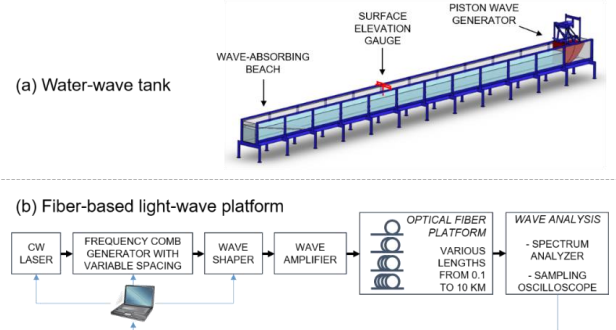

FIG. 5. Experimental setups for nonlinear propagation of dn- and cn-periodic (a) water waves and (b) light waves.

In water wave experiments conducted in deep-water conditions, the initial periodic wave profiles are shaped with a piston wave generator located at one end of the tank. An electric signal drives the piston to directly modulate the surface height in the time domain according to the exact mathematical expression for the surface elevation. The tank dimensions are $30 \times 1 \times 1$ m$^3$ and the water depth is 0.7 m. A wave-absorbing beach is installed at the opposite end to avoid the influence of reflected waves. The effective wave propagation in the absence of wave reflection in the flume is 20 m, considering the wave maker and beach installations in this set-up. Seven wave gauges are then placed at distinct distances from the wave excitation to record the discretized evolution of surface elevation in the longitudinal direction of wave propagation. Given the upper frequency range limit of the wave maker that can generate waves with a frequency up to 2 Hz only, which in deep-water correspond to a wavelength of 0.4 m following the dispersion relation, the effective wave propagation distance corresponds to 50 wavelengths and is therefore sufficient to observe respective breather growth dynamics for a wide range of wave steepness values. Nevertheless, this propagation distance is still insufficient to observe extreme wave formations from pure noise excitation. A key to successful experiments is to accurately generate the exact surface elevation scales in the boundary conditions, as described by the theory, and to choose the carrier parameters (in particular the carrier steepness) properly to avoid physical wave breaking during the wave focusing process [34].

*Experimental results.* We first performed experimental measurements of the spontaneous MI gain when periodic waves propagate in km-long optical fibers. Two distinct 60- and 40-GHz frequency combs are initially generated to subsequently shape exact periodic solutions (respectively for dn- and cn-waves) according to usual values of fiber dispersion $\beta_2$ (-21 ps$^2$ km$^{-1}$) and nonlinearity $\gamma$ (1.2 W$^{-1}$ km$^{-1}$), and the optical peak power $A_0$ suitably chosen and injected. The fiber loss $\alpha$ is very low about 5% per kilometer (in power). The correspondence between theory and experiment is the following: dimensional distance $z$ (m) and time $t$ (s) are given by $z = \xi L_{NL}$ and $t = \tau t_0$, where the characteristic length and time scales are $L_{NL} = (\gamma A_0)^{-1}$ and $t_0 = (|\beta_2| L_{NL})^{\frac{1}{2}}$, respectively. The dimensional optical field $E$ (W$^{1/2}$) is $E = A_0^{\frac{1}{2}} \psi$.

Figure 6(a-b) shows the recorded power spectra $|\tilde{E}|^2$ and the accumulated MI gain obtained for various propagation distances in the case of a dn-periodic wave ($k = 0.7$). We clearly observe the MI gain bands emerging around the central peak of the dn-wave in Fig. 6(a) and growing exponentially with propagation distance. Their bandwidth remains limited by the intrinsic frequency spacing of the comb formed by the dn-wave (i.e., 60 GHz). The limited resolution of the spectrum analyzer however prevents from measuring accurately the full MI bands in the vicinity of the comb peaks. This explains why the accumulated MI gain was obtained over a limited range of frequency detuning in Fig. 6(b). Even so these results are in good agreement with theoretical predictions of MI gain (the accumulated power gain over a normalized distance $\Delta \xi$ is obtained as $20 \log_{10}[e^{\text{Re}\{\Gamma\}\Delta\xi} \sin(\text{Im}\{\Gamma\}\Delta\xi)]$). This confirms that we exceed a 20-dB maximal gain after 5 km of fiber at a frequency detuning of two-thirds of the comb frequency interval. The corresponding normalized distances for the theory were calculated from effective fiber lengths (this includes a partial correction of fiber losses to the nonlinear propagation distance through $L_{eff} = [1 - e^{-\alpha L_{fiber}}]/\alpha$ [4]). We can note the effect of fiber losses on the power of each comb harmonic after a few kilometers in Fig. 6(a). This effect is typically accompanied by a strong decrease of the modulation contrast of the dn-wave in the time domain, and even some phase-shift pulsations [24].

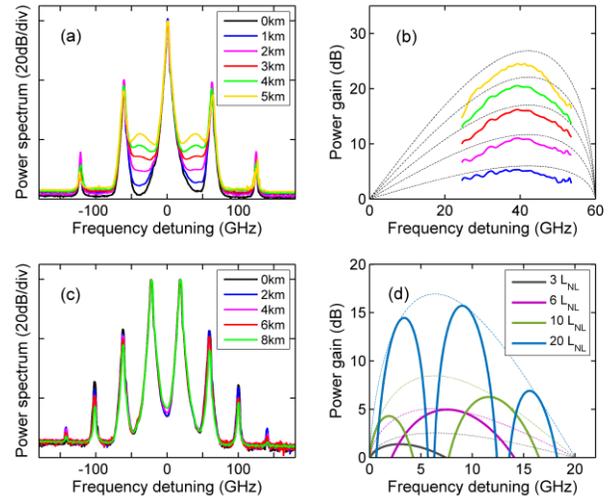

FIG. 6. (a) Experimental power spectra of spontaneous MI for a dn-wave ($k = 0.7$) on various propagation distances (here $A_0 = 0.83$ W). (b) Accumulated MI power gain deduced from (a) and compared to theory (black dotted lines) calculated from respective normalized distances 0.97, 1.88, 2.74, 3.56 and 4.33 $L_{NL}$ based on the effective fiber lengths. (c) Experimental power spectra of spontaneous MI for a cn-wave ($k = 0.92$) on various distances (here $A_0 = 0.63$ W). No MI gain is clearly observed as the corresponding normalized distances are respectively 1.43, 2.70, 3.84, and 4.85 $L_{NL}$ based on the effective fiber lengths. (d) Theoretical predictions of accumulated MI power gain, for the cn-wave studied in (c), when longer distances are considered. Solid (dotted) lines are calculated with (without) Im$\{\Gamma\}$.

For the case of a cn-periodic wave, the results are depicted in Fig. 6(c). We investigated the evolution of the power spectrum over 8 km of propagation and no clear signature of MI gain was observed, except a few-dB gain around the center frequency (i.e., zero-detuning). Different values of the modulus $k$ were studied with similar results. The apparent robustness of cn- waves needs to be moderated since the studied normalized distance is only 4.85 $L_{NL}$ after correction of fiber losses. In addition, two main issues could be raised based on Fig. 6(c), namely the limited resolution of the spectrum analyzer that prevents from better distinguishing MI gain bands from the comb formed by the cn-wave, and the significant impact of fiber losses again observed on the power of each comb harmonic. But, in any case, Fig. 6(d) confirms that MI gain bands would be readily observable only if longer normalized distances are considered beyond 10 $L_{NL}$ (i.e., a 13-km-long fiber without any propagation loss).

We emphasize here that the predictions are calculated by taking into account the nonzero imaginary part of $\Gamma$, thus describing oscillations of the MI gain with distance. More specifically, it traduces the frequency location for the fastest growing perturbation after a certain distance with a maximum defined by Re$\{\Gamma\}$. We clearly show that the accumulated MI gain differs in bandwidth and shapes for distinct propagation distances. In the first steps of propagation, a first spectral band emerges and then continuously drifts to larger frequency detunings. By increasing the propagation distance, extra sub-bands appear and fill the frequency interval defined by Re$\{\Gamma\}$ (dotted lines) while the overall gain also increases. Such complex behaviours were confirmed by numerical simulations of NLSE (see Fig. 3), but their direct observation appears as a hard task from the experimental point of view.

In addition to the spontaneous MI, we carried out specific experiments in both optics and hydrodynamics about the coherent seeding of the process, and more particularly on the generation of rogue breathers (RW solutions) on stationary periodic backgrounds. To this end, we use the exact solutions (6) and (7) to shape the input periodic wave with the correct localized perturbation. According to the maximal propagation distance that can be reached, we chose suitable initial conditions ($\xi$ value) to observe the maximum amplification.

Figure 7(a,c) presents both temporal and spectral evolutions measured for the RW on the dn-periodic wave (4) with the fiber-based light-wave platform. Our light shaping technique implies the time-periodic generation of localized perturbations, so that a 13-GHz frequency comb was initially generated to subsequently shape $\psi_{\text{dn}}^{\text{RW}}(\tau, \xi = -2.3)$ at fiber input. Note that the frequency interval for the dn-periodic wave is 78 GHz. In Fig. 7(a) we clearly reveal that the localized perturbation (centered at $t = 0$) grows as predicted by the theory shown in Fig. 7(b). A typical X-wave interaction forms in the space-time plane. The optical RW reaches a maximum amplitude nearly 3 W$^{\frac{1}{2}}$ after 1.4 km, which is close to the theoretical prediction $A_0^{\frac{1}{2}}(2 + \sqrt{1-k^2})$ despite the fiber losses. As next, we also observe the RW decay just before 2 km. As expected, the nonlinear focusing of perturbation induces a significant spectral broadening shown in Fig. 7(c) and satisfies the corresponding theory from Fig. 7(d).

We now describe the experiments performed in the water-wave tank. We recall that the surface elevation $\eta(z, t)$ is related to the NLSE wave envelope $\psi(z, t)$ up to second order in steepness by using the computational formula: $\eta(z,t) = \text{Re}\left\{\Psi(z,t)e^{i(\beta z-\omega t)} + \frac{1}{2}\kappa\,\Psi^2 e^{2i(\beta z-\omega t)}\right\}$. The correspondence between theory and experiment can be retrieved here by using the fact that $a\Psi(a^2 z, at)$ is also a solution and the following relations hold: $\xi = \kappa z$, $\tau = \omega\left(t - z/c_g\right)/\sqrt{2}$ and $\psi = \frac{\Psi}{a}$, where $a$ and $\kappa$ are the initial amplitude and the wave number of the carrier wave, respectively. These two parameters define the steepness $a\kappa$, whereas the dispersion relation of linear deep-water wave theory gives the angular frequency $\omega = (g\kappa)^{\frac{1}{2}}$, where $g$ is the gravitational acceleration. The group velocity is equal to $c_g = \omega/2\kappa$. The carrier amplitude has chosen to be $a = 0.01$ m and its frequency 1.7 Hz for all tests. This corresponds to a wave frequency of $\omega = 10.68$ s$^{-1}$ and a wavenumber $\kappa = 11.63$ m$^{-1}$. The attenuation rate in our water-wave experiments was estimated about 0.25% per meter (in amplitude), which means that the experienced dissipation (see also [29]) for RW generation will be larger here than in the optical experiment reported in Fig. 7.

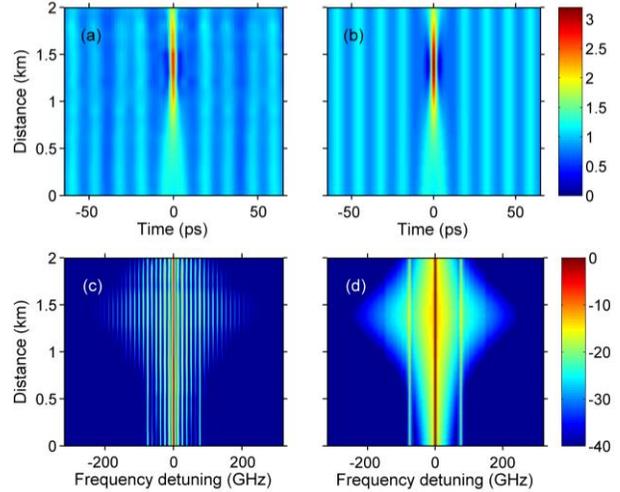

FIG. 7. Results for the RW on the dn-periodic wave ($k = 0.7$) with light-waves. (a-b) Longitudinal evolution of the optical envelope $|E(z,t)|$ obtained from experiment and theory, respectively (here $A_0 = 1.38$ W). (c-d) Corresponding evolution of power spectrum $|\tilde{E}|^2$ (in log. scale, dB unit) from experiment and theory, respectively.

Figure 8 shows the results of experiments by shaping an initial localized perturbation centered at $t = 0$ onto the dn- and cn-periodic waves for $\xi = -2.6$. The two first cases (Fig. 8a-d) report the longitudinal evolution of perturbation for dn-periodic waves when $k = 0.8$ and 0.99 (i.e., close to the soliton limit) until reaching the maximal amplification after 16.8 m. In both cases, the measurements agree well with theory. For $k = 0.8$, the overall picture is very similar to the one reported in optics (see Fig. 7(a)), while for $k = 0.99$ the periodic background wave is weaker and the dynamics clearly approaches a two-soliton interaction with a maximal $M$ factor near 2 as is predicted by the theory. The two last cases shown in Fig. 8(e-h) correspond to the case of cn-periodic waves. Again, the experimental results are in accordance with the theory. For $k = 0.95$, we retrieve similar dynamics as just previously mentioned since we are close to the soliton limit for cn-waves. Now when changing $k$ to 0.75, one can notice the significant decrease of the nonlinear focusing experienced by the perturbation, and more particularly we confirm that maximal amplitude of the RW structure is $2k$ according to the theory. Additional experiments with distinct values of $k$ confirmed that the amplification factor is always close to 2 (independent of $k$).

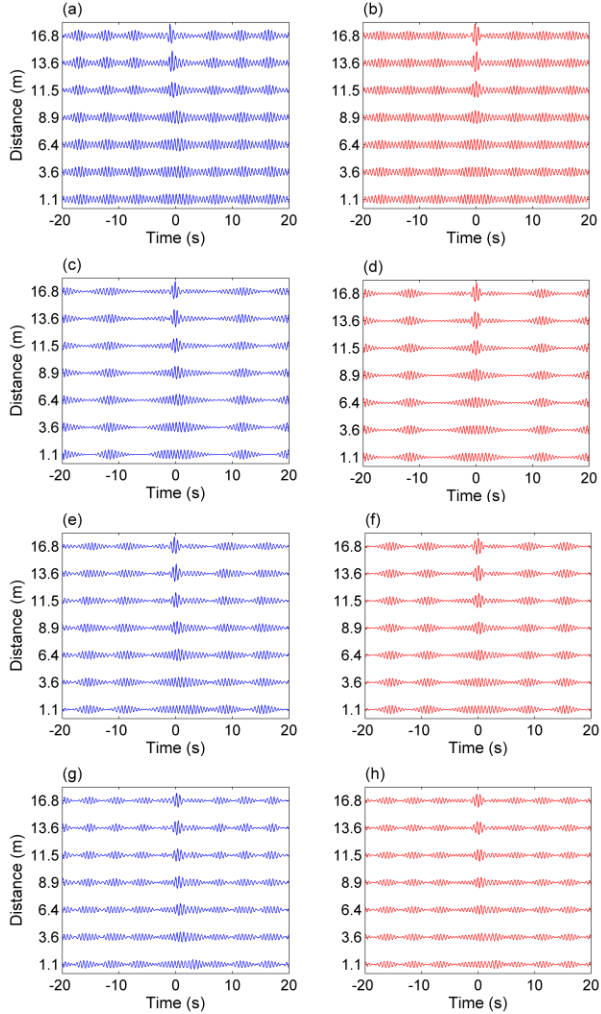

FIG. 8. Results for RWs with water waves ($a = 0.01$ m, and $a\kappa = 0.116$). Left panels: Evolution of time series of surface elevation measurement with propagation distance. Right panels: Corresponding theory. (a-b) Rogue dn-periodic wave ($k = 0.8$). (c-d) Rogue dn-periodic wave ($k = 0.99$). (e-f) Rogue cn-periodic wave ($k = 0.95$). (g-h) Rogue cn-periodic wave ($k = 0.75$).

In fact, the maximal extreme water surface measured and predicted, as shown in Fig. 8, namely the cases {(a),(b);(c),(d);(e),(f);(g),(h)} in meters are {0.027,0.027;0.022,0.023;0.022,0.019;0.016,0.014}, respectively. Note that in idealized experimental conditions with negligible dissipation, such as the case when the wave channels are much wider compared to current set-up, the amplification rate could be even higher, depending on the significant local steepness reached in the focused wave group. Consequently, the higher-order Stokes contributions become more significant by changing the shape of the wave by making the troughs flatter and crests sharper, compared to purely linear sinusoidal form. In addition, higher-order effects, summarized in higher-order dispersion and mean flow contributions, are responsible for the asymmetry of the wave group. These can be accounted for in the modified NLSE, also known as the Dysthe equation [11].

Due to the limited number of wave gauges, which is of seven, it is not possible to display the evolution as accurately as in Fig. 7, which is the optical counterpart, or as in Ref. [29]. One possibility to overcome this limitation is to repeat the same experiment again and again while changing the positions of the gauges to increase the spatial resolution. However, we are motivated to show the results as measured from a single run while wave profiles are measured with respect to this exact latter test. The comparison with theory clearly confirms the extreme wave growth on the respective periodic envelope profiles.

*Conclusion.* In summary, we reported the theoretical description and direct observation of the MI process and related rogue breathers on stationary periodic dnoidal and cnoidal envelopes. The present work was performed in two distinct disciplines of wave physics, namely, optics and hydrodynamics, in order to confirm the existence of the MI phenomenon for more complex background waves than the common plane wave. We provided an overview of the main characteristics of MI gain and RWs for various values of the modulus of elliptic functions.

Future experimental research should for instance tackle the complexity of MI gain for cn-periodic waves and the instability of other elliptic wave solutions such as the double-periodic solutions [35-36]. We also expect that our multidisciplinary approach will motivate new scientific and technological advances in the field of nonlinear physics, telecommunications and marine engineering.

*Acknowledgments.* AC acknowledges David Kedziora and Nail Akhmediev for useful discussion. DP thanks Robert White for collaboration. BK acknowledges financial support of the French "Investissements d'Avenir" program (PIA2/ ISITE-BFC, ANR-15-IDEX-03, "Breathing Light" project). Theoretical work was supported by the Russian Science Foundation (No.19-12-00253).